\documentclass[a4paper,fleqn,usenatbib]{mnras}

\usepackage{txfonts}

\usepackage[T1]{fontenc}
\usepackage{ae,aecompl}

\usepackage{psfig}   
\usepackage{graphicx}
\usepackage{amssymb}
\usepackage{subfigure}

\title[Making BEASTies]{Making BEASTies: dynamical formation of planetary systems around massive stars}

\author[R. J. Parker \& E. C. Daffern-Powell]{
Richard J. Parker\thanks{E-mail: R.Parker@sheffield.ac.uk}\thanks{Royal Society Dorothy Hodgkin Fellow} and Emma C. Daffern-Powell
\\
Department of Physics and Astronomy, The University of Sheffield, Hicks Building, Hounsfield Road, Sheffield S3 7RH, UK}

\pubyear{2022}

\begin{document}
\label{firstpage}
\pagerange{\pageref{firstpage}--\pageref{lastpage}}
\maketitle

\begin{abstract}
 Exoplanets display incredible diversity, from planetary system architectures around Sun-like stars that are very different to our Solar System, to planets orbiting post-main sequence stars or stellar remnants. Recently the B-star Exoplanet Abundance STudy (BEAST) reported the discovery of at least two super-Jovian planets orbiting massive stars in the Sco~Cen OB association. Whilst such massive stars do have Keplerian discs, it is hard to envisage gas giant planets being able to form in such hostile environments. We use $N$-body simulations of star-forming regions to show that these systems can instead form from the capture of a free-floating planet, or the direct theft of a planet from one star to another, more massive star. We find that this occurs on average once in the first 10\,Myr of an association's evolution, and that the semimajor axes of the hitherto confirmed BEAST planets (290 and 556\,au) are more consistent with capture than theft. Our results lend further credence to the notion that planets on more distant ($>100$\,au) orbits may not be orbiting their parent star.   

\end{abstract}





\section{Introduction}

Planetary systems are ubiquitous in the the Galaxy, with more than 5000 discovered to date\footnote{https://exoplanetarchive.ipac.caltech.edu/docs/counts\_detail.html}, and many more candidate systems requiring follow-up observations and confirmation. Furthermore, the planetary systems observed around other stars are incredibly diverse, from systems of tightly-packed, close-in terrestrial planets \citep{Borucki13} to Hot Jupiters \citep{Mayor95}, to super-Jupiter ($\gtrsim$~10\,M$_{\rm Jup}$) mass planets often orbiting at large distances from their host stars \citep{Marois08,Quanz10}. 

The latter group of planets, almost all found via high-contrast imaging, are particularly interesting as they perhaps more than most challenge our ideas of planet formation. Whilst they may form \emph{in situ} through disc fragmentation \citep{Boss97,Mayer02}, they may have formed closer in to their host stars and subsequently moved via dynamical processes, either with other planets in the system \citep{Davies13}, or through the direct \citep{Laughlin98,Smith01,Parker12a} and indirect \citep{Fabrycky07,Malmberg07a,Parker09c} influence of other stars in the stellar birth environment. 

Furthermore, recent work has shown that planets can be captured (where they are free-floating in the star-forming region, and then become bound to another star following an encounter), or even stolen in a direct exchange between two stars, with the planet ending up on a bound orbit around the other star \citep{DaffernPowell22}.

In addition to solar- and low-mass stars (around which planetary systems are ubiquitous in the Galactic disc), star-forming regions also host more massive ($\gtrsim 2.4$\,M$_\odot$) O- and B-type stars, which due to their mass, and the fact they are often in binary systems \citep{Sana13,Villasenor21}, have a higher cross section for encounters and thus may be able to steal or capture planets more efficiently than low-mass stars.  

Recently, \citet{Janson21} initiated the B-star Exoplanet Abundance STudy (BEAST) to detect giant planets orbiting massive OB type stars, stars with masses $\gtrsim$2.4\,M$_\odot$. The study immediately bore fruit with the detection of a super-Jupiter-mass planet orbiting at 556\,au around $b$~Cen~AB, a 6--10\,M$_\odot$ binary star \citep{Janson21b}; and two similar-mass planets orbiting at 21\,au  and 290\,au around $\mu^2$~Sco, a 9\,M$_\odot$ star \citep{Squicciarini22}.  

Whilst massive stars are observed to host Keplerian discs \citep{Cesaroni05,Johnston15}, which could in principle facilitate the rapid formation of gas giant planets, the intense radiation fields emitted from these stars are likely to cause photoevaporation of the disc, which will significantly hinder, or even prevent planet formation \citep[e.g.][]{Armitage00,Nicholson19a}. In this Letter we explore an alternative scenario for making the BEAST systems, namely that they are stolen or captured in their birth star-forming regions. The BEAST systems (hereafter ``BEASTies'') are observed in OB associations, which typically have a low global stellar density \citep{Wright14}, but have pockets of dense substructure in which dynamical encounters (and the theft and capture of planetary mass objects) can occur \citep{Parker14b}. Alternatively, some authors posit that OB associations were much more dense at formation, and that stellar feedback mechanisms cause the rapid expansion of these regions \citep{Kroupa01a}. Both of these scenarios are consistent with the simulations we adopt in this work.

The Letter is organised as follows. We outline our methods in Section~\ref{sec:Methods}, we present our results in Section~\ref{sec:ResultsDiscussion} and we draw conclusions in Section~\ref{sec:conclusions}.

\section{Method}
\label{sec:Methods}

We use a subset of the $N$-body simulations described in \citet{DaffernPowell22}, which contain $N_\star = 1000$ stars, drawn from a \citet{Maschberger13} IMF with a probability distribution of the form
\begin{equation}
p(m) \propto \left(\frac{m}{\mu}\right)^{-\alpha}\left(1 + \left(\frac{m}{\mu}\right)^{1 - \alpha}\right)^{-\beta}.
\label{maschberger_imf}
\end{equation}
Here, $\mu = 0.2$\,M$_\odot$ is the scale parameter, or `peak' of the IMF \citep{Bastian10,Maschberger13}, $\alpha = 2.3$ is the \citet{Salpeter55} power-law exponent for higher mass stars, and $\beta = 1.4$ describes the slope of the IMF for low-mass objects \citep*[which also deviates from the log-normal form;][]{Bastian10}. We randomly sample this distribution in the mass range 0.1 -- 50\,M$_\odot$, such that brown dwarfs are not included in the simulations. This distribution is sampled stochastically, so different realisations of the same simulation contain different numbers of massive stars, but we obtain between 44 and 65 stars with masses $\gtrsim 2.4$\,M$_\odot$, the lower-mass limit for host stars in the BEAST papers.

For simplicity (and to reduce computational expense) we do not include primordial stellar binaries, although these are ubiquitous in star-forming regions \citep{Duchene13b}, and the first BEASTie discovered orbits a massive star binary. The effect of ignoring binaries is to effectively reduce the numbers of stolen and captured stars in the simulations, because a binary presents a larger cross section for interaction. Half of the stars with masses $<$2.4\,M$_\odot$ are randomly assigned a 1\,M$_{\rm Jup}$ planet with semimajor axis $a_p = 30$\,au and zero eccentricity.

The stars (and their planetary systems) are distributed within a box-fractal distribution \citep{Goodwin04a,DaffernPowell20} to mimic the spatial and kinematic substructure observed in many star-forming regions. We adopt a fractal dimension $D=1.6$, which is the highest degree of substructure possible in three dimensions. The velocities are set such that nearby stars have similar velocities (i.e\,\,a small local velocity dispersion), whereas distant stars can have very different velocities, similar to the observed \citet{Larson81} laws. This high degree of substructure facilitates interactions early on in the simulations, even if the density decreases due to the dynamical expansion of the regions.
We set the radius of the fractals to be $r_F = 1$\,pc, resulting in an \emph{initial} median local stellar density in the fractals of $\tilde{\rho} \sim 10^4$M$_\odot$\,pc$^{-3}$. 

We scale the velocities of the stars such that the global virial ratio is  $\alpha_{\rm vir} = T/|\Omega|$, where $T$ and $|\Omega|$ are the total kinetic and potential energies, respectively. We adopt an initial virial ratio $\alpha_{\rm vir} = 1.5$, which is an unbound, supervirial velocity field. The kinematic substructure in the fractals is set up in such a way that the local clumps of stars may be bound, or even subvirial, but the overall motion of the star-forming region is to expand, due to the virial ratio. 

These initial conditions are adopted as a guess at the initial conditions of the Sco Cen OB association targeted by BEAST. \citet{Kroupa01a} postulated that OB associations are the expanded remnants of compact star clusters, whereas other authors have shown that the kinematics of these regions suggests that they were never more dense in the past \citep{Wright14,Wright16,Ward18}. More recent work, however, has shown evidence of expansion in other regions \citep{Kounkel18,Quintana22}, which has been attributed to feedback.

Whilst our simulations do not include stellar feedback, they expand overall (which could be due to feedback, or the region could simply have formed unbound) whilst still containing substructure akin the the filamentary appearance of young star-forming regions.

Although our simulations are designed to mimic a Sco~Cen-like association, it is not our aim to exactly match the observed region in detail. The Sco Cen region is comprised of three sub groups, and also borders another star-forming region, Ophiuchus \citep{Preibisch08}. At best, our simulations are only a very rough approximation of the region.

To assess the statistical significance, we run 20 realisations of the same simulation, identical apart from the random number seed used to initialise the initial mass, velocity and position distributions. The simulations are evolved for 10\,Myr using the \texttt{kira} integrator within the \texttt{Starlab} environment \citep{Zwart99,Zwart01}. We do not include stellar evolution in the simulations, although mass-loss from massive stars could change the orbits of planets once they have been captured or stolen.

We analyse the simulations at an age of 10\,Myr. In \citet{DaffernPowell22}, the rate of theft and capture of planets was highest earlier on (between 0.1 -- 2\,Myr), but many of these planets were not stable in the long-term. At 10\,Myr, our simulations have evolved past the stage at which captured/stolen planets could be further disrupted by stellar encounters. We therefore have a better idea of which stolen/captured planets will be stable in the long-term, and we also are comparing them to ages commensurate with the association targeted by the BEAST observations \citep[the age of Sco~Cen is likely between 10--20\,Myr,][]{Pecaut16,Janson21}.

\begin{figure*}
  \begin{center}
\setlength{\subfigcapskip}{10pt}
\hspace*{-1.3cm}\subfigure[]{\label{fig:associations-a}\rotatebox{270}{\includegraphics[scale=0.24]{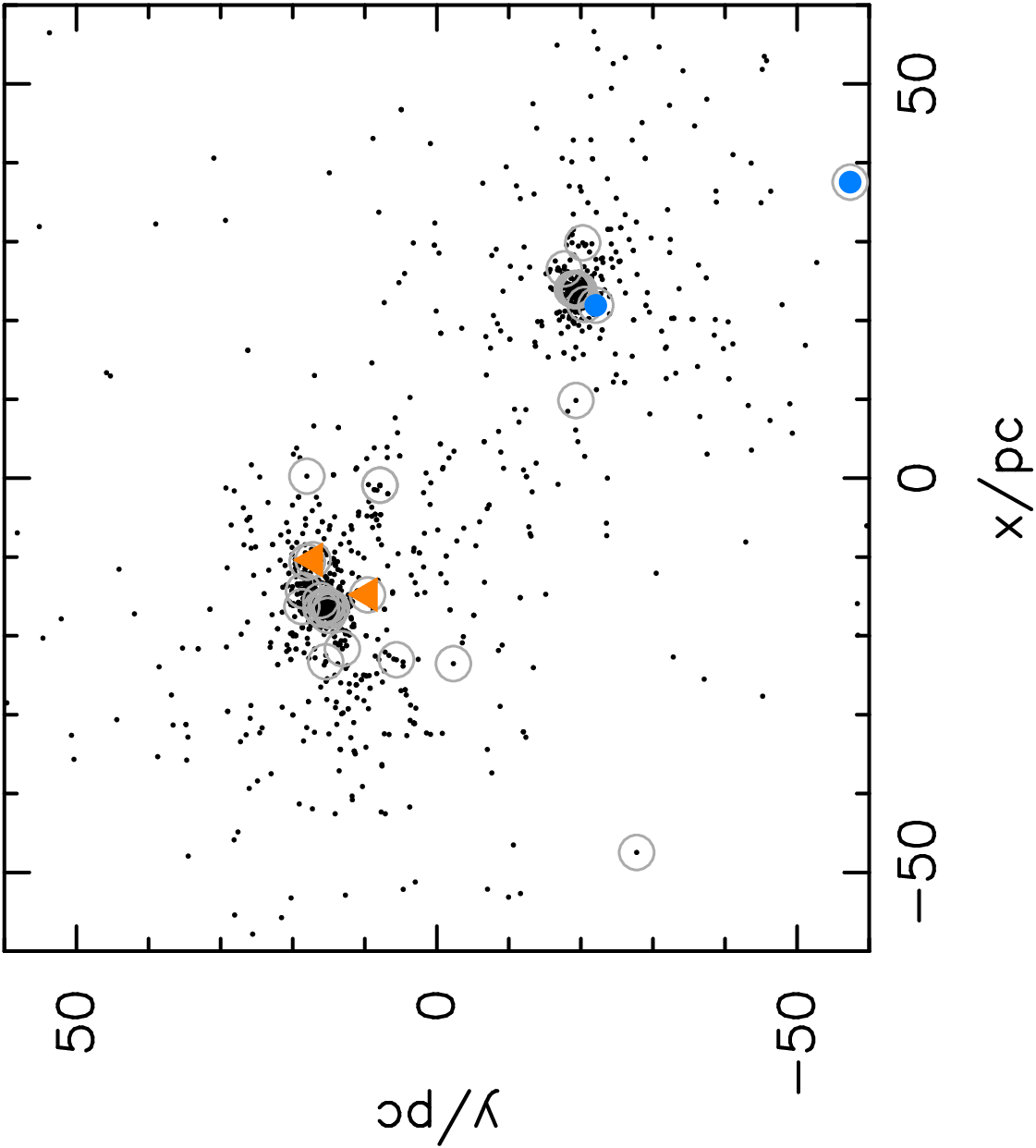}}}
\hspace*{0.3cm}\subfigure[]{\label{fig:associations-b}\rotatebox{270}{\includegraphics[scale=0.24]{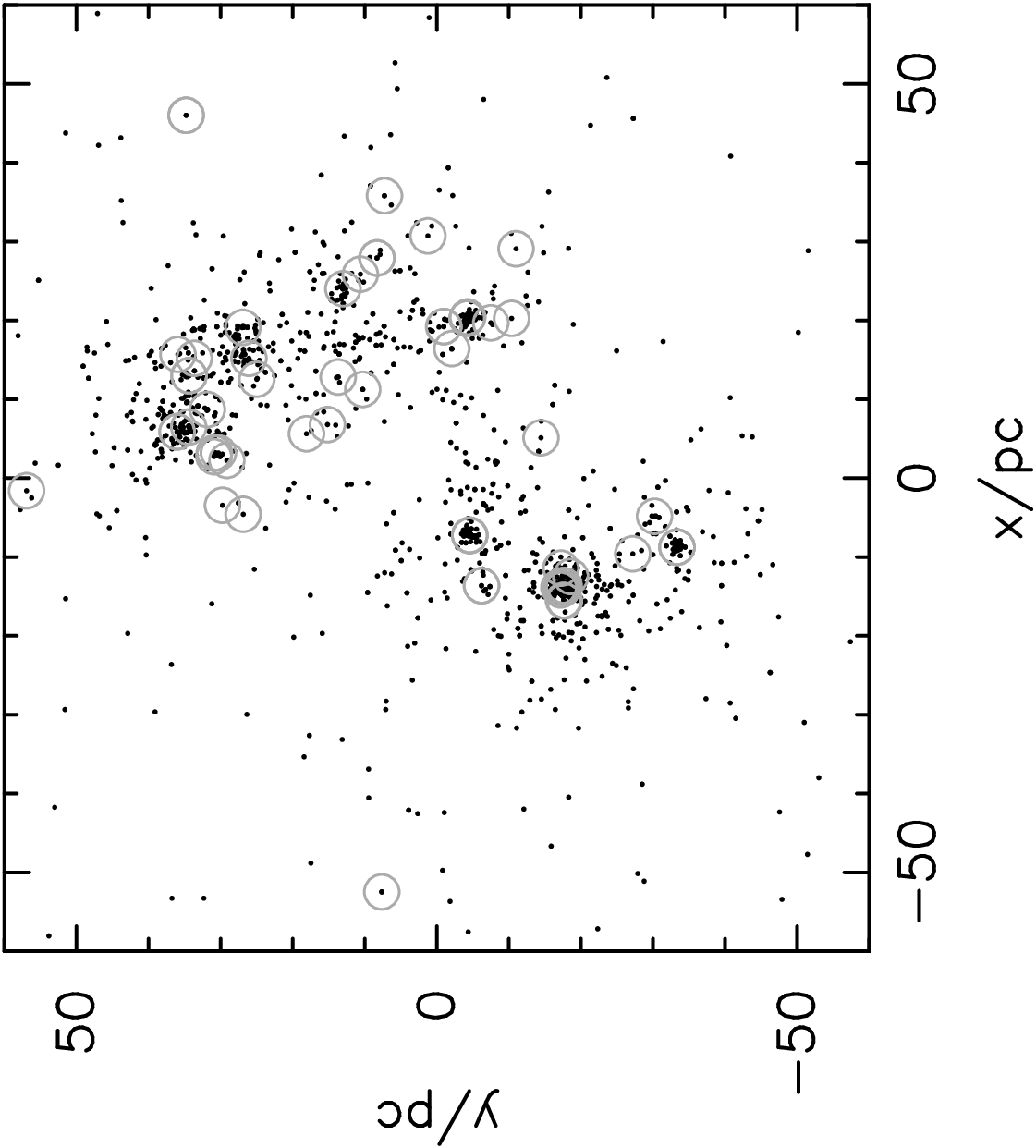}}}
\hspace*{0.3cm}\subfigure[]{\label{fig:associations-c}\rotatebox{270}{\includegraphics[scale=0.24]{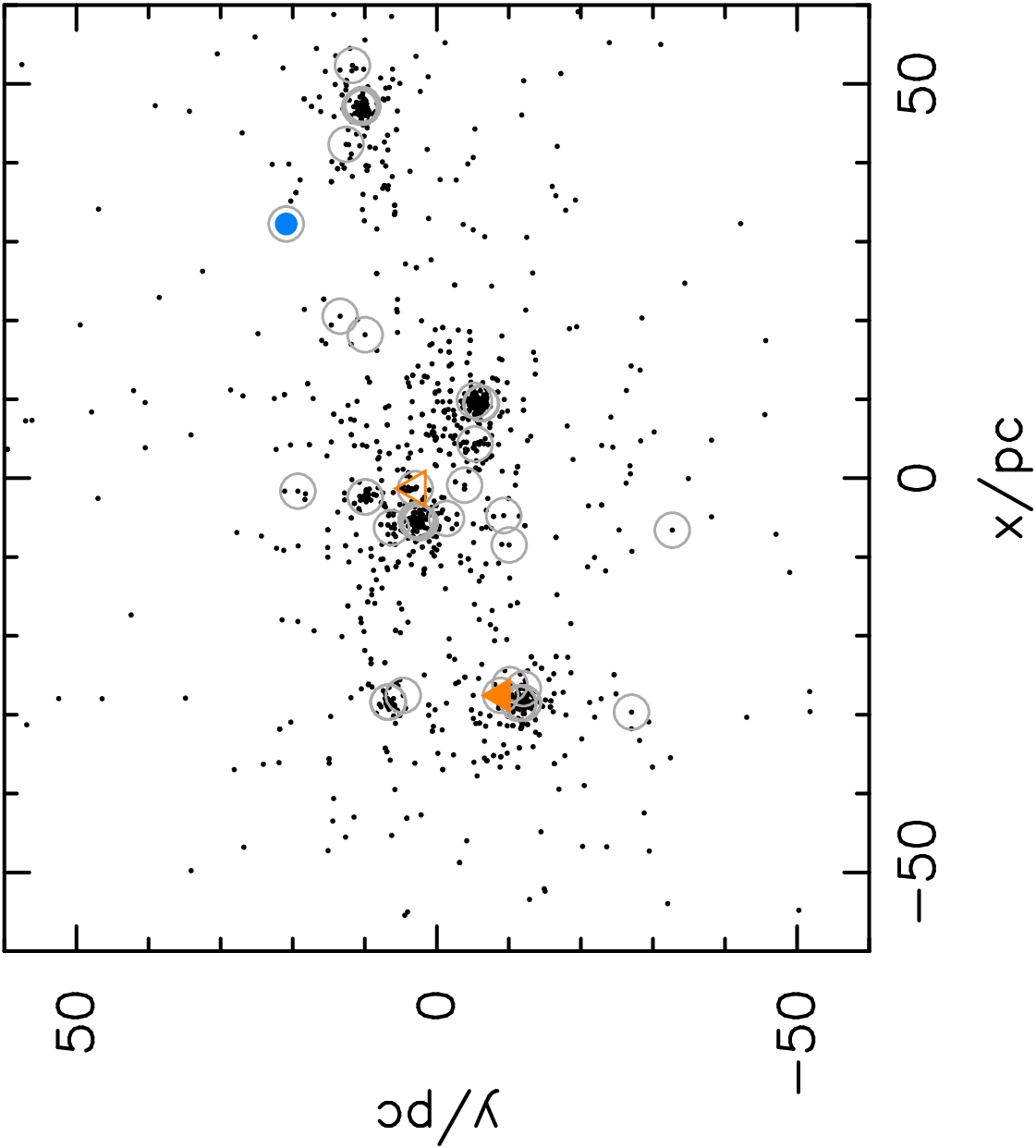}}}
\caption[bf]{Snapshots of three simulated OB associations after 10\,Myr of evolution. On the left is a binary cluster, and centre and right show more filamentary-like morphologies. In panels (a) and (c) the filled orange triangles and blue circles are captured or stolen BEASTie planets, respectively, whereas the open orange triangle in panel (c) is a BEASTie within a triple system. In all panels, OB stars are indicated by the grey circles, and we note that in panel (b) this association contains no BEASTies.}
\label{fig:associations}
  \end{center}
\end{figure*}

\section{Results}
\label{sec:ResultsDiscussion}

We search for stolen planets -- those that are directly exchanged between stars and do not spend any time as unbound free-floating planets, as well as captured planets -- which do spend some time as a free-floating planet before forming a bound orbit around another star. We then categorise the planets based on the host star's mass; if the star mass is $\geq$2.4\,M$_\odot$, we classify the planet as a BEASTie.

In Fig.~\ref{fig:associations}, we show three examples (out of twenty) of the spatial distributions of the simulated OB associations after 10\,Myr. These simulations sometimes evolve into binary clusters \citep{Arnold17}, i.e.\,\,two subclusters  of stars orbiting each other (Fig.~\ref{fig:associations-a}). In around 60\,per cent of the simulations, the spatial confiuration is more filamentary (e.g.~Figs.~\ref{fig:associations-b}~and~\ref{fig:associations-c}).

In two of the three panels in Fig.~\ref{fig:associations}, the OB associations form BEASTies, shown by the solid orange triangles (captured) and solid blue circles (stolen). Planets that form part of a triple system are shown by the orange open triangle, and all the OB stars are shown by the grey circles. We find no correlation between the numbers of BEASTies and the morphology of the assoication, or on the overal numbers of OB stars. 

As the number of stars $\geq$2.4\,M$_\odot$ is inherently small due to the nature of the initial mass function (between 44 and 65, depending on the simulation, compared to around 85 targets in the BEAST observations, which are of a more populous region than our simulations), and the overall combined frequency of captured and stolen planets is only $\sim$4\,percent \citep{DaffernPowell22}, we do not necessarily expect many (if any) BEASTies in each simulation.

Across twenty simulations (identical aside from the random number seed used to initialise the masses, positions and velocities of the stars), we find 7 stolen planets, and 11 captured planets orbiting massive stars. In other words, we expect around one BEASTie per star-forming region, although in some instances we may have two or three BEASTies in one region, and none in another. We therefore posit that not every observed OB association should contain BEASTies; if observations subsequently show these systems are more common, or are found in \emph{every} star-forming region, this would suggest an alternative formation mechanism for these objects.

\begin{figure}
    \centering
    \rotatebox{270}{\includegraphics[scale=0.35]{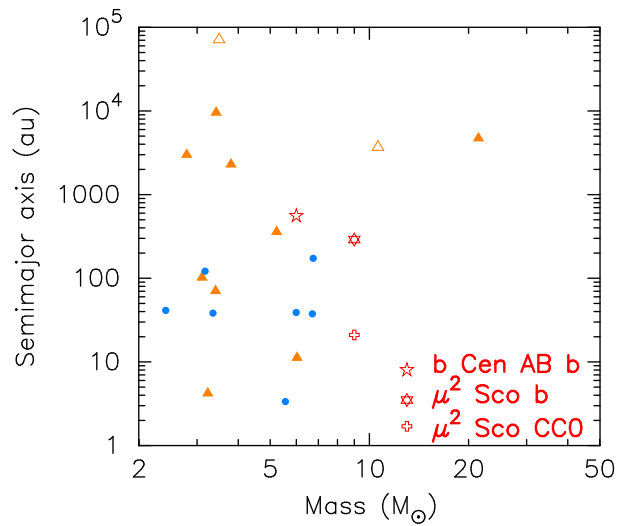}}
    \caption{Semimajor axis versus host star mass of captured (orange triangles) and stolen (blue circles) BEASTies in all twenty of our simulations. Planets that are part of a triple system are shown by the open orange triangles. We also show the observed objects from the BEAST project by the red points \citep{Janson21b,Squicciarini22}.}
    \label{fig:sma_mass}
\end{figure}

In Fig.~\ref{fig:sma_mass} we show the semimajor axis versus mass of the BEASTies in our simulations, as well as the three observed examples from \citet{Janson21b} and \citet{Squicciarini22}. The captured BEASTies in our simulations are shown by the orange triangles, and the stolen BEASTies are shown by the blue circles.   The most `extreme' system formed in our simulation is when one of the captured BEASTies orbits a 21\,M$_\odot$ star at a semimajor axis of 4727\,au. We do not include stellar evolution in our simulations, but at the age at which we perform our analysis (10\,Myr) we would probably expect this star to have already exploded as a supernova, or at the very least have traversed the giant branch \citep[e.g.][]{Limongi06}.

We find two instances of planets which are part of a triple system that form via capture, shown by the open triangles in Fig.~\ref{fig:sma_mass}. One shows a triple system consisting of a planet orbiting a 0.15\,M$_\odot$ star, which forms a triple with a much more distant 3.5\,M$_\odot$ star. The semimajor axis of the outer orbit is 71376\,au, with a very high eccentricity of 0.97. This system forms late in the simulation (at 9.5\,Myr), but is seen in each of the subsequent snapshots. The second system is similar, in that a low-mass (0.20\,M$_\odot$) star with a planet orbiting at 27\,au forms a triple with a 10.6\,M$_\odot$ star, where the outer orbit semimajor axis is 3700\,au and eccentricity $e = 0.80$. This system forms much earlier, at 1\,Myr, and is present throughout. 

\begin{figure}
    \centering
    \rotatebox{270}{\includegraphics[scale=0.35]{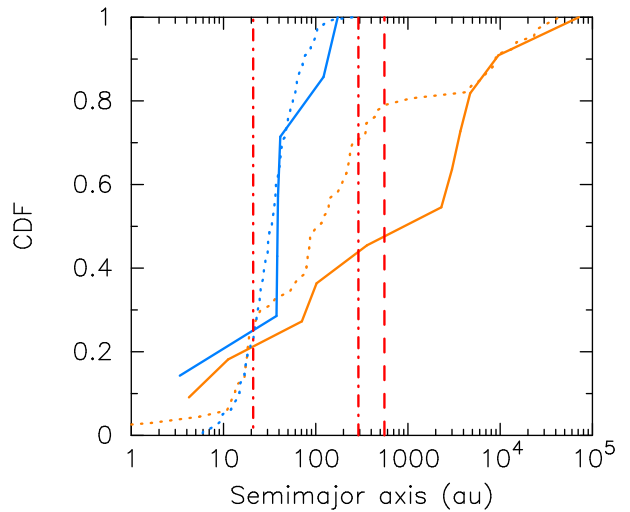}}
    \caption{Semimajor axes distributions of stolen (solid blue curve) and captured (solid orange curve) BEASTies in all twenty simulations. For comparison, we show the semimajor axes of planets stolen or captured by low-mass ($<$2.4\,M$_\odot$) stars by the dotted blue and orange curves, respectively. For comparison, we show the semimajor axes for the observed BEAST systems by the vertical red lines; the planets around $\mu^2$~Sco are shown by the dot-dashed lines and the planet orbiting $b$~Cen~AB is shown by the dashed line. }
    \label{fig:sma_dists}
\end{figure}

We find that several of our BEASTies lie close to the values for the three observed systems in the BEAST papers. In Fig.~\ref{fig:sma_dists} we show the cumulative distributions of semimajor axes of stolen and captured planets around stars $m \geq 2.4$\,M$_\odot$ (the solid blue and orange curves, respectively), as well as the distributions for planets stolen or captured around stars with mass $m<2.4$\,M$_\odot$ (the dotted blue and orange curves, respectively).

For comparison, we show the semimajor axes for the observed BEAST systems by the vertical red lines; the planets around $\mu^2$~Sco are shown by the dot-dashed lines and the planet orbiting $b$~Cen~AB is shown by the dashed line. We note that no stolen planets in our simulations have semimajor axes $>$200\,au, so we posit that the $b$~Cen~AB~b and $\mu^2$~Sco~b planets (semimajor axes 556\,au and 290\,au, respectively) likely formed via capture, with theft more likely for the $\mu^2$~Sco~CC0 (candidate) planet.

From inspection, the semimajor axes distributions of the stolen BEASTies are similar to the semimajor axes of stolen planets around low-mass stars (compare the dashed blue and solid blue curves in Fig.~\ref{fig:sma_dists}), but the distributions for captured planets (the dotted and solid orange curves) appear visually different. However, from a KS test we cannot reject the hypothesis that they share the same underlying parent distribution.

\begin{figure}
    \centering
    \rotatebox{270}{\includegraphics[scale=0.35]{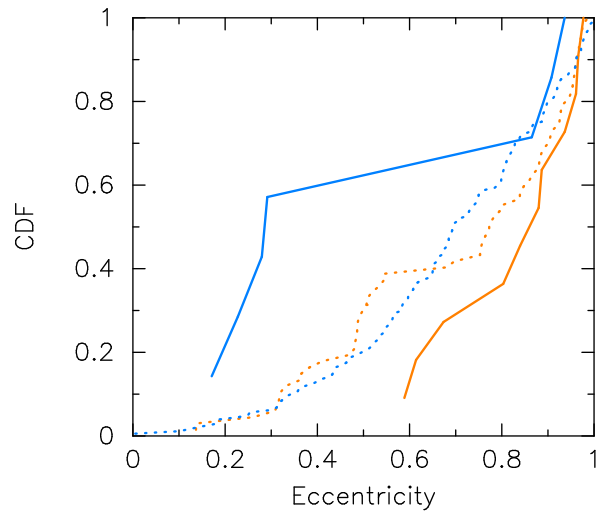}}
    \caption{Orbital eccentricity distributions of stolen (solid blue curve) and captured (solid orange curve) BEASTies in all twenty simulations. For comparison, we show the semimajor axes of planets stolen or captured by low-mass ($<$2.4\,M$_\odot$) stars by the dotted blue and orange curves, respectively.}
    \label{fig:ecc_dists}
\end{figure}

In Fig.~\ref{fig:ecc_dists} we show the cumulative distributions of orbital eccentricities of the BEASTies  formed in our simulations through theft or capture by stars $m \geq 2.4$\,M$_\odot$ (the solid blue and orange curves, respectively). We also show the distributions for planets stolen or captured around stars with mass $m<2.4$\,M$_\odot$ (the dotted blue and orange curves, respectively).

When performing a KS-test on the distributions, the low p-value ($<0.1$) might lead us to reject the hypothesis that the stolen BEASTies share the same underlying parent eccentricity distribution as those forming around lower mass stars, despite their formation channels being identical. The eccentricity distributions of the captured BEASTies, and the stolen and captured planets around lower-mass stars, are consistent with being drawn from a thermal distribution \citep{Heggie75}, which is expected for binary systems that form dynamically \citep[see also][]{Perets12}. 

\section{Conclusions}
\label{sec:conclusions}

Motivated by the BEAST study \citep{Janson21} We perform $N$-body simulations of star-forming regions that evolve into OB associations and determine the number of planets that are captured or stolen onto orbits around OB stars ($m \geq 2.4$\,M$_\odot$), as well as their orbital properties. Each simulation contains 1000 stars, with a planet orbiting around  half of the low-mass ($<2.4$\,M$_\odot$) stars. Our conclusions are the following:

(i) Across 20 simulations each containing $\sim$450 planets, a total of 11 planets are captured by OB stars, and a further 7 are stolen from a lower mass star  during 10\,Myr of dynamical evolution. 

(ii) The semimajor axes of the captured BEASTies range between $4$\,au -- $10^5$\,au, whereas the stolen planets span a narrower range (3\,au -- 200\,au). As two of the three planets discovered so far by BEAST orbit at distances $>$200\,au, we posit that these planets are captured as their host OB association dynamcially evolves.

(iii) The masses of the OB stars are usually between 2 -- 8\,M$_\odot$, suggesting that if the planets remain dynamically stable, they would not immediately be affected by the star exiting the main sequence. We do find one system, however, that consists of a planet orbiting a 21\,M$_\odot$ star at several thousand au.

(iv) The semimajor axis distributions of stolen and captured BEASTies are similar to those of planets formed the same way around low-mass stars. Intriguingly, the eccentricity distributions are similar for captured planets, but different for stolen planets, although we are hamstrung by low-number statistics.

(v) There is an average of just less than one BEASTie planet per simulation, although this is stochastic; some associations may form up to four BEASTies, whereas other associations do not form any.  

Our results are an extension of previous work that shows planets at distances $>$100\,au may not be orbiting their parent star \citep{Parker12a,Perets12,Li15,Mustill16,DaffernPowell22}.

\section*{Acknowledgements} 
We thank the anonymous referee for a prompt and helpful report. RJP acknowledges support from the Royal Society in the form of a Dorothy Hodgkin Fellowship. 

\section*{Data Availability}
The data underlying this article will be shared on reasonable request to the corresponding author.




\bibliographystyle{mnras}
\bibliography{general_ref} 




\bsp	
\label{lastpage}
\end{document}